\documentclass[prl,aps,showpacs,twocolumn,preprintnumbers,nofootinbib,amsmath,amssymb,superscriptaddress,longbibliography]{revtex4-1}

\usepackage{systeme}
\usepackage{mathtools}
\usepackage[inline]{enumitem}
\usepackage{graphicx}
\usepackage{upgreek}

\usepackage[T1]{fontenc}
\usepackage[english]{babel}
\usepackage{xcolor}
\usepackage{braket}
\usepackage{nicefrac}
\usepackage{bm}
\usepackage{bbm}
\usepackage{braket}
\usepackage{bbold}    
\usepackage{slashed}
\usepackage[normalem]{ulem}
\usepackage{array}
\newcolumntype{C}{>{$}c<{$}}

\renewcommand{\dag}{\dagger}

\newcommand{\beqn}{\begin{eqnarray}}
\newcommand{\eeqn}{\end{eqnarray}}
\newcommand{\beqs}{\begin{subequations}}
\newcommand{\eeqs}{\end{subequations}\\[-2mm]\noindent}
\newcommand{\eq}[1]{(\ref{#1})}

\newcommand{\bs}{\boldsymbol}
\newcommand{\avr}[1]{{\left\langle #1 \right\rangle}}

\newcommand{\ave}[1]{\langle #1 \rangle}

\allowdisplaybreaks

\definecolor{brickred}{rgb}{0.8, 0.25, 0.33}
\definecolor{macouleur}{RGB}{105,150,150}

\usepackage{color}
\definecolor{purple}{rgb}{0.8,0,0.6}

\begin{document}

\title{Boundary states and Non-Abelian Casimir effect in lattice Yang-Mills theory}

\author{Maxim N. Chernodub}
\affiliation{Institut Denis Poisson UMR 7013, Universit\'e de Tours, 37200 Tours, France}

\author{Vladimir A. Goy}
\affiliation{Pacific Quantum Center, Far Eastern Federal University, 690950 Vladivostok, Russia}

\author{Alexander V. Molochkov}
\affiliation{Pacific Quantum Center, Far Eastern Federal University, 690950 Vladivostok, Russia}

\author{Alexey S. Tanashkin}
\affiliation{Pacific Quantum Center, Far Eastern Federal University, 690950 Vladivostok, Russia}

\begin{abstract}
Using first-principle numerical simulations, we investigate the Casimir effect in zero-temperature SU(3) lattice gauge theory in 3+1 spacetime dimensions. The Casimir interaction between perfect chromometallic mirrors reveals the presence of a new gluonic state with the mass $m_{\mathrm{gt}} = 1.0(1)\sqrt{\sigma} = 0.49(5)\,\mathrm{GeV} = 0.29(3) M_{0^{++}}$ which is substantially lighter than the $0^{++}$ groundstate glueball. We call this excitation ``glueton'' interpreting it as a non-perturbative colorless state of gluons bound to their negatively colored images in the chromometallic mirror. The glueton is a gluonic counterpart of a surface electron-hole exciton in semiconductors. We also show that a heavy quark is attracted to the neutral chromometallic mirror, thus supporting the existence of a ``quarkiton'' (a ``quark exciton'') colorless state in QCD, which is formed by a single quark with its anti-quark image in the chromometallic mirror. Analogies with edge modes in topological insulators and boundary states of fractional vortices in multi-component condensates are highlighted.
\end{abstract}

\date{\today}

\maketitle

\paragraph{\bf Introduction.}

The presence of physical macroscopic objects affects fluctuations of quantum fields in the vacuum around them while the modified fluctuations, in turn, exert the Casimir-Polder force on these physical objects~\cite{Casimir1948}. The phenomenon, known as the Casimir effect~\cite{casimir1948attraction}, is often interpreted as the experimentally measurable evidence~\cite{Lamoreaux1997,Mohideen1998,Bressi2002} of the vacuum energy associated with the ``zero-point'' quantum fluctuations~\cite{Milton2001,Bordag2009}. This interpretation originates from the fact that vacuum fluctuations influence neutral objects, such as perfectly conducting neutral metallic objects, that carry no electric or other charges with their dipole and higher-order moments vanishing. Another interpretation of the Casimir effect is given in Ref.~\cite{Jaffe2005}.

The Casimir energy depends not only on the geometry of the objects but also on interactions of the quantum fields~\cite{Milton2001,Bordag2009}. However, in a phenomenologically relevant case of Quantum Electrodynamics, a correction to the tree\--level Casimir effect due to electron-photon interactions is so tiny that it cannot be observed with the existing experimental technology~\cite{Bordag1985}.

Remarkably, in strongly coupled theories, the effects of boundaries are much more pronounced: they modify not only the vacuum forces but also influence the structure of the vacuum itself. For example, analytical studies of effective models suggest the existence of Casimir-induced phase transitions in fermionic effective field theories~\cite{Flachi_2013bc,Flachi_2017cdo} and the $\mathbb{C}P^{N-1}$ model on a finite interval~\cite{Flachi_2017xat,Betti_2017zcm}. In addition, first-principle numerical simulations show that boundaries in interacting gauge theories, such as compact electrodynamics~\cite{Chernodub_2017mhi,Chernodub_2022izt} and Yang-Mills theory in two spatial dimensions~\cite{Chernodub_2018pmt}, affects nonperturbative properties, including mass gap generation and (de)confinement (for a review see Ref.~\cite{Chernodub_2019nct}).

Moreover, bounded systems often possess new degrees of freedom that emerge exclusively due to the presence of boundaries. These boundary states and associated boundary central charges ignite substantial interest in the conformal field theories~\cite{Herzog_2017xha,Andrei_2018die}. They also appear in the condensed matter systems as the celebrated edge states in topological insulators~\cite{Konig_2008} that have deep roots in the lattice field theory~\cite{Kaplan_1992bt}.

In our paper, we aim to uncover new, nonperturbative boundary states in Yang-Mills theory and put a bridge between the two phenomena, the Casimir effect and the edge states in the scope of phenomenologically relevant SU(3) Yang-Mills theory in 3+1 dimensions. What is the relation between the restructuring of the gluonic vacuum in bounded geometries~\cite{Chernodub_2018pmt} -- related, in particular, to phenomenologically relevant MIT bag model~\cite{Chodos_1974je,Chodos_1974pn} -- and possible boundary states in Yang-Mills theory? To this end, we first address the Casimir effect on the lattice.

\vskip 1mm
\paragraph{\bf Lattice Casimir setup.}
The formulation of the Casimir problem in lattice gauge theories have been first discussed for Abelian gauge theories in Refs.~\cite{Pavlovsky_2009mt,Pavlovsky_2009kg,Chernodub_2016owp} with the extension to the investigation of nonperturbative features of Abelian~\cite{Chernodub_2017mhi,Chernodub_2022izt} and non-Abelian~\cite{Chernodub_2018pmt} lattice gauge theories and, more recently, to free fermionic lattice models~\cite{Ishikawa:2020ezm,Ishikawa:2020icy,Nakayama:2022fvh,Mandlecha:2022cll}. Below, we will briefly remind the essential points of the construction for gauge theories referring the interested reader for more details to Ref.~\cite{Chernodub_2016owp}.

\begin{figure}[!thb]
\centering
  \includegraphics[width=0.975\linewidth]{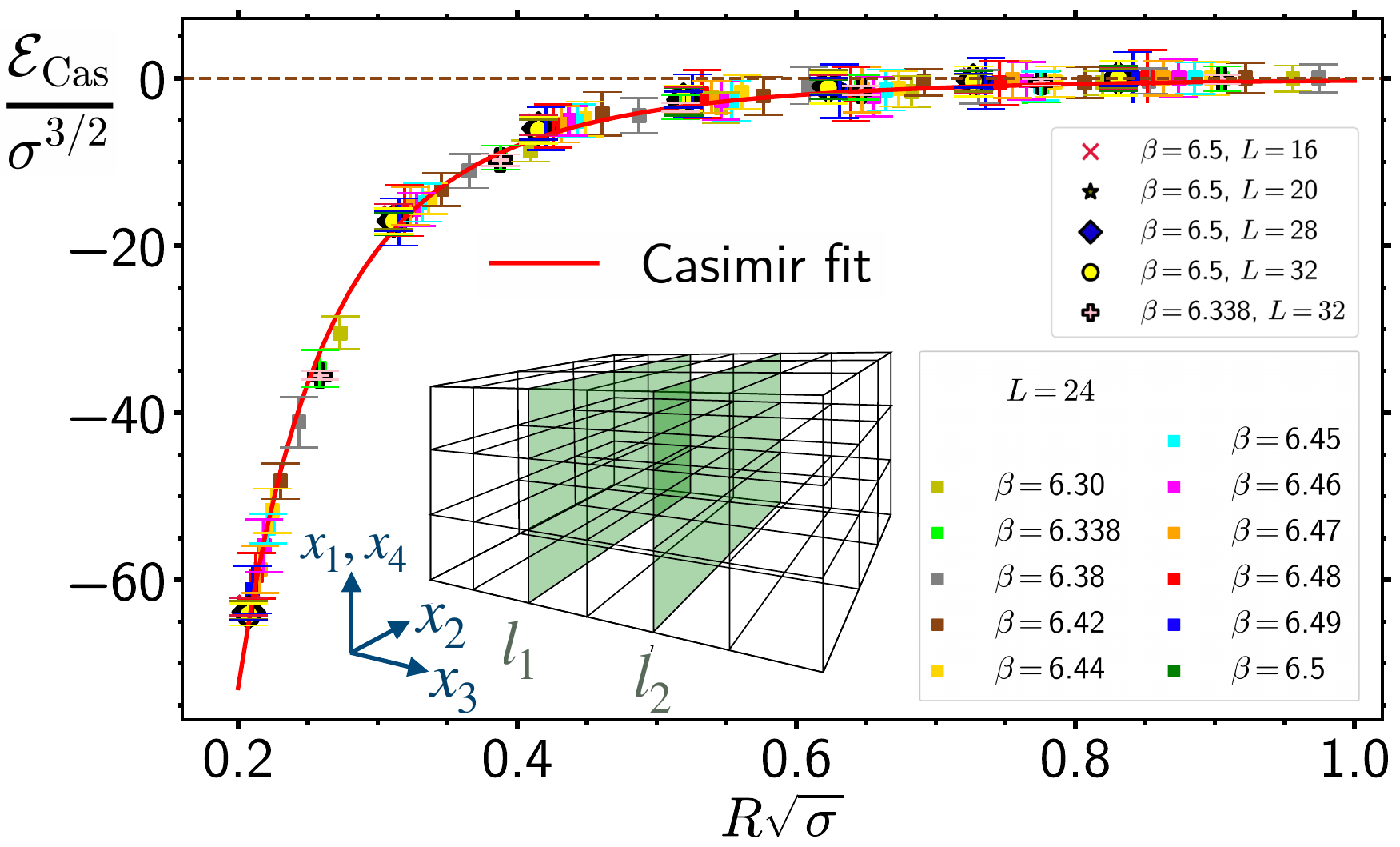}
  \caption{Casimir energy density ${\mathcal E}_{\mathrm{Cas}}$ vs. distance $R$ between the perfect chromometallic plates in units of the fundamental string tension $\sigma$ for various values of the lattice coupling $\beta$ and several lattice volumes $L^4$. The line shows the best fit by the phenomenological function~\eq{eq_Casimir_fit}, representing the Casimir energy of a massive field. The inset illustrates the Casimir double-plate geometry on the lattice with $R = |l_2 - l_1|$.}
  \label{fig_Casimir_energy}
\end{figure}

The non-Abelian Casimir setup features two perfectly conducting flat chromometallic plates in the $(x_1,x_2)$ plane separated by the distance $R = |l_1 - l_2|$ along the $x_3$ axis, as it is shown in the inset of Fig.~\ref{fig_Casimir_energy}. For Yang-Mills theory with the action in $(3+1)d$ Minkowski spacetime,
\begin{equation}
    S = - \frac{1}{4} \int d^4 x \, F_{\mu\nu}^a F^{a,\mu\nu} \,.
    \label{eq_S_continuum}
\end{equation}
the gauge-invariant Casimir boundary conditions are:
\begin{equation}
	E^a_{\|}(x) {\biggl|}_{x\in S} \! = B^a_{\perp}(x) {\biggl|}_{x\in S} \! = 0, 
    \qquad
    a = 1,\dots, N^2 -1.
    \label{eq_Casimir_continuum}
\end{equation} 
They imply that the tangential chromoelectric $E^a_i \equiv F^a_{0i}$ and normal chromomagnetic fields $B^a_i = (1/2) \varepsilon_{ijk} F^{a,jk}$ vanish at the surface $S$ (in our case, $S$ is the set of two planes). Conditions~\eq{eq_Casimir_continuum} are identical, up to the color index $a = 1,\dots, N^2 -1$, to the conditions imposed on the Abelian electromagnetic (photon) field at the surface of a perfectly conducting metal (a mirror) in electrodynamics. Thus, Eqs.~\eq{eq_Casimir_continuum} correspond to a chromometallic mirror plate for gluons.

The Wilson form of the lattice Yang-Mills action is given by a sum over lattice plaquettes $P {\equiv} P_{n,\mu\nu} {=} \{n,\mu\nu\}$:
\beqn
		S = \beta\sum_{P} \left( 1 - {\mathcal P}_P \right), \qquad
      {\mathcal P}_P = \frac{1}{3}\mathrm{Re\,Tr}\,U_P,\quad
\label{eq_action}
\eeqn
where $\mu$ and $\nu$ label directions, $n$ denotes a site of a $4d$ Euclidean lattice, and $\beta = 6/g^2$ is the lattice coupling. In continuum limit, the lattice spacing vanishes, $a\to 0$, the lattice plaquette $U_{\mu\nu}(n) = U_\mu(n)U_\nu(n+\hat\mu)U^\dag_\nu(n+\hat\nu) U^\dag_\nu(n) = \exp(i a^2 F_{\mu\nu}(n) + \mathcal{O}(a^3))$ reduces to the continuum field-strength tensor~$F_{\mu\nu}$, and the lattice action~\eq{eq_action} becomes a Euclidean version of Yang-Mills action~\eq{eq_S_continuum}. 

The Casimir boundary conditions~\eq{eq_Casimir_continuum} in the Euclidean lattice formulation are achieved by promoting the lattice coupling in Eq.~\eq{eq_action} to a plaquette-dependent quantity, $\beta \to \beta_P$, where $\beta_P = \lambda \beta$ if the plaquette $P$ either touches or belongs to the hypersurface spanned by the surface $S$ and $\beta_P = \beta$ otherwise~\cite{Chernodub_2016owp}. The quantity $\lambda$ plays a role of a Lagrange multiplier which, in the limit $\lambda \to \infty$, enforces the lattice version of Eqs.~\eq{eq_Casimir_continuum}. 

In Minkowski spacetime, the canonical energy-momentum tensor reads as follows: 
\beqn
T^{\mu\nu} = F^{\mu\alpha} F^{\nu}_{\ \alpha} - \frac{1}{4} \eta^{\mu\nu} F^{\alpha\beta} F_{\alpha\beta}
\eeqn
where $\eta_{\mu\nu} = {\mathrm{diag}}\,(1,-1,-1,-1)$ is the metric. The energy density ${\mathcal E} $ is related to its Euclidean counterpart as
\beqn
{\mathcal E} \equiv T^{00} = \frac{1}{2} \left( {\bs B}^2 + {\bs E}^2\right) \to T^{44}_E = \frac{1}{2} \left( {\bs B}^2_E - {\bs E}^2_E \right),\ 
\label{eq_E_Wick}
\eeqn
where the superscript ``$E$'' labels the Euclidean quantities. The lattice Casimir energy density per unit area of the Casimir plates on the zero-temperature lattice of the volume $L_s^4$ is given by the properly normalized lattice version of Eq.~\eq{eq_E_Wick}:
\begin{equation}
    {\mathcal E}_{\mathrm{Cas}} = \beta L_s \biggl(\,\sum_{i=1}^3\ave{{\mathcal P}_{i4}}_S 
    - \sum_{i<j = 1}^3 \ave{{\mathcal P}_{ij}}_S \biggr),
\label{eq_Cas_Lattice}
\end{equation}
where average plaquettes are taken over the whole lattice volume. Quantity~\eq{eq_Cas_Lattice} represents the difference between the vacuum expectation values of temporal and spacial plaquettes in the presence of the mirror plates $S$. An additive divergent contribution to the expectation values of plaquettes, arising from zero-point ultraviolet fluctuations, cancels exactly in Eq.~\eq{eq_Cas_Lattice}. The Casimir energy density~\eq{eq_Cas_Lattice} is a finite physical quantity that depends only on the distance between $R$ the mirrors and vanishes in their absence (or, at $R \to \infty$). 

We perform simulations at zero temperature $L_s^4$ lattices of various volumes, $L_s = 12, 16, 20, 28, 32$ using 13 values of the gauge coupling varying in the range from $\beta = 5.6924$ to $\beta = 6.5$. The physical scaling of the lattice spacing $a = a(\beta)$ is set via the phenomenological value of the fundamental string tension, $\sqrt{\sigma} = 485(6)\,\mathrm{MeV} = [0.407(5)\,\mathrm{fm}]^{-1}$ following Ref.~\cite{Athenodorou_2020ani}. Values of $a\sqrt{\sigma}$ for intermediate $\beta$'s, which cannot be found in Ref.~\cite{Athenodorou_2020ani}, are obtained from an accurate spline interpolation. To generate and update gauge field configurations, we used the Monte-Carlo heatbath algorithm~\cite{Gattringer2010,Boudreau2018}. For each point, set by the gauge coupling constant $\beta$ and the lattice distance between plates $R/a$, we generated $6 \times 10^5$ trajectories. The first $10^5$ configurations are omitted to achieve thermalization. Next, we proceed to the numerical calculation of the Casimir energy~\eq{eq_Cas_Lattice} on the lattice.

\vskip 1mm
\paragraph{\bf Non-Abelian Casimir energy and the glueton.}

Figure~\ref{fig_Casimir_energy} shows the Casimir energy density between the chromometallic plates~\eq{eq_Cas_Lattice} calculated from first principles in SU(3) gauge theory. The data for a broad set of lattice volumes and couplings nicely collapse to a smooth curve, thus demonstrating the absence of substantial finite-size and finite-volume effects. 

The Casimir energy takes a large negative value as the inter-plate separation $R$ diminishes. This behavior points to the attractive nature of the non-Abelian Casimir force expected at short separations, where gluons should experience the asymptotic freedom and the Casimir interactions should reduce to the one of a free massless vector field with a color degeneracy factor. 

At large inter-plate separations $R$, the Casimir energy expectedly vanishes. In theories with a free massless field, the Casimir energy density per unit plate area drops as an inverse power $R^{-3}$ of the distance $R$, while in field theories with a mass $m \neq 0$, one expects that the Casimir energy density vanishes exponentially, ${\mathcal E}(R) \sim e^{- 2 m R}$. The factor $2$ implies that the particle has to travel from one mirror plate to another, then get reflected to close the path, thus propagating the distance $2R$ in total. Therefore, it is crucial to determine how rapidly the energy diminishes in the large--$R$ limit, as this behavior should uncover the mass spectrum of excitations in the gluonic vacuum between the chromometallic mirrors. 

In (2+1) dimensional confining theories, closely spaced chromometallic boundaries are known to affect the vacuum structure between them~\cite{Chernodub_2017gwe,Chernodub_2022izt}. In SU(2) Yang-Mills theory, the lowest excitation between the plates corresponds to a ``Casimir particle'' with a mass substantially lower than the lowest glueball mass in the same theory~\cite{Chernodub_2018pmt}. The Casimir mass is related to the magnetic mass in 2+1 dimensional Yang-Mills theory~\cite{Karabali_2018ael}. 

The nonperturbative Casimir energy in $(2+1)d$ non-Abelian gauge theory can successfully be described as the Casimir energy of a massive scalar particle~\cite{Karabali_2018ael}. Applying the same idea in $(3+1)$ dimensions, we fit our numerical results with the Casimir energy of a scalar field~\cite{Hays_1979bc,Ambjorn_1981xw,Cougo-Pinto_1994obn} with certain mass $m_{\mathrm{gt}}$:
\begin{equation}
    {\mathcal E}_{\mathrm{Cas}} = - C_0 \frac{2 (N_c^2 - 1) m^2_{\mathrm{gt}}}{8 \pi^2 R}\sum_{n=1}^{\infty}\frac{K_2(2n m_{\mathrm{gt}} R)}{n^2}\,.
    \label{eq_Casimir_fit}
\end{equation}
The prefactor takes into account the $(N_c^2 - 1)$-fold color degeneracy (with $N_c = 3$ in our case) as well as two spin polarization of (massless) gluons. The mass gap could affect this factor, thus forcing us to include a phenomenological parameter $C_0$. The sum in Eq.~\eq{eq_Casimir_fit} is performed over a quickly converging series of modified Bessel functions of the second kind~$K_2(x)$. 

The best fit of the Casimir energy by function~\eq{eq_Casimir_fit} is shown in Fig.~\ref{fig_Casimir_energy} by the red line. The fit (with $\chi^2/{\mathrm{d.o.f.}} \simeq 0.6$ highlighting its good quality) provides us with the following best-fit parameters: $C_0 = 5.60(7)$ and 
\begin{align}
    m_{\mathrm{gt}} = 1.0(1)\sqrt{\sigma} = 0.49(5)\,\mathrm{GeV}\,.
    \label{eq_mass_glueton}
\end{align}

Strikingly, the mass of the exchange particle~\eq{eq_mass_glueton} is substantially smaller than the mass of the groundstate glueball $M_{0^{++}} = 3.405(21) \sqrt{\sigma} = 1.653(26) \, \mathrm{GeV}$~\cite{Athenodorou_2020ani}. Moreover, the result~\eq{eq_mass_glueton} is surprising because the groundstate glueball mass $M_{0^{++}}$, by its very definition, is identified with the lowest possible mass in the system. The same phenomenon has been found for an effective particle that governs the long-distance limit of the Casimir effect in two spatial dimensions~\cite{Chernodub_2018pmt}. Nevertheless, we found an excitation with the nonzero mass~\eq{eq_mass_glueton}, which is substantially lower than the lowest groundstate mass.

The apparent contradiction is resolved by noticing that the groundstate glueball mass $M_{0^{++}}$ determines the mass gap in {\it the bulk} of the system (far from eventual boundaries) while the mass~\eq{eq_mass_glueton} is associated with a new excitation in Yang-Mills theory that emerges exclusively due to the presence of {\it a boundary}. We call this boundary state ``glueton'' interpreting it as a nonperturbative colorless state of gluons bound to their negatively colored images in the chromometallic mirror. 

The states localized at the boundaries of a system (often called the edge states) can have lower masses than the mass gap in the bulk of the same system. In the condensed matter context, this effect appears at the contacts of semiconductor structures (the Volkov-Pankratov states~\cite{volkov1985two}) and the boundaries of topological insulators (massless edge modes featuring the spin Hall effect~\cite{hasan2010colloquium,qi2011topological}). However, contrary to the mentioned boundary modes, the glueton has a non-topological origin.

The glueton is a non-Abelian analog of a surface exciton that emerges in electronic systems. The surface exciton is an electrically neutral quasiparticle that exists in semiconductors and insulators close to their boundaries: an electron (or a hole) in the bulk of the material couples to its image hole (electron) state in the reflective boundary and forms a neutral quasiparticle~\cite{mills1982surface}. These electron-hole states can only move along the boundary of the material. The physics of surface excitons constitutes a vast area of research in solid-state physics~\cite{Cocoletzi_2005,Agranovich2009,gavrilenko2011optics,dean2012excitons}.

The glueton should be distinguished from another gluonic excitation, so-called ``gluelump''~\cite{Michael1985,Campbell_1985kp,Jorysz_1987qj,Philipsen1999}. The gluelump is a purely gluonic system consisting of a valence gluon connected by an adjoint string to a static adjoint source which can be associated with an infinitely heavy gluon. Although the gluelump is not a physical object that cannot be directly measured in an experiment, its theoretical investigation provides valuable insight into the nonperturbative confining properties of QCD~\cite{Simonov_2000ky}. Furthermore, contrary to the gluelump, the glueton can propagate along a reflective domain wall in QCD (for example, along the vacuum-hadronic interface in an MIT bag model~\cite{Chodos_1974je,Chodos_1974pn}) and thus can potentially contribute to the stability of such states and associated physically measurable quantities. 

For completeness of our description, we also mention that Yang-Mills theory possesses yet another, ``torelon'', excitation which appears in systems with a compact spatial dimension~\cite{Lucini_2001nv}. The torelon corresponds to a confining flux tube that winds around a spatial torus and has no fixed color sources. It has a numerically calculable spectrum corresponding to the eigenstates of the stretched confining string, which cannot collapse due to geometrical topological reasons~\cite{Juge_2003vw,Athenodorou_2017cmw}. Recently it has been revealed that the ground state of the torelon corresponds to an axion-type excitation on the world sheet of the closed flux tube~\cite{Athenodorou_2017cmw,Athenodorou_2021vkw}. 
 
The glueton (a surface state) is yet another gluonic excitation in addition to the glueball (a bulk state), the gluelump (a heavy-light gluon-bound state), and the torelon (a stretched-string state). 

\vskip 1mm
\paragraph{\bf The quarkiton: a quark bound by a mirror.}

We argued above that the chromometallic Casimir plate, acting as a mirror for gluons, facilitates the creation of a colorless (glueton) state bound of gluons to their mirror images. One can question whether a quark can form a colorless bound state with its negative image in a chromometallic mirror, a ``quarkiton''?  

This ``quark--chromometallic mirror'' bound state is expected to be strengthened by the color confinement phenomenon. Indeed, in the bulk of the confinement phase, the chromoelectric field of a quark is squeezed into the confining string, which terminates, in a meson, on an anti-quark. If we place a quark near the non-Abelian mirror, the confining string should terminate on the mirror, thus attracting the quark to its negative image. Therefore, we expect to observe the confinement of a quark with a neutral chromometallic mirror via the formation of the confining QCD string. 

Since the mirror is a globally color-neutral object, the induced color charge, which mimics the image anti-quark at the mirror, should lead to a re-distribution of the color charge over the surface of an (infinite) mirror. In a confining system, the redistributed charge can contribute positively to the total free energy of the quark-mirror system, and it can, in principle, outweigh the negative contribution of the quarkiton bound state.

As we study a purely gluonic system, we cannot check the formation of the quarkiton state by calculating the mass spectrum with quark degrees of freedom near the mirror. However, we can calculate the potential $V_{Q|}$ (which is given by the free energy $ V_{Q|} \equiv F_{Q|}(d)$) produced by the mirror ``$|$'' on the heavy quark ``$Q$'' separated by the distance $d$. Moreover, the slope of the potential allows us to estimate whether (and how strongly) the quark is attracted to (or repelled by) the mirror. 

Associating the potential of the static quark with its free energy $F_{Q|}(d)$, we use the Polyakov loop operator, which places a static heavy quark at the spatial point $\bs x$:
\begin{align}
    P_{\bs x} = \frac{1}{3} {\mathrm{Re}}\, {\mathrm{Tr}} \left( \prod_{x_4=0}^{L_t - 1} U_{{\bs x},x_4} \right)\,,
    \label{eq_Polyakov_loop}
\end{align}
where the product over the time-like oriented non-Abelian $U_{{\bs x},x_4}$ matrices is closed via the periodic boundary conditions. The effect of the boundary mirror is identified via the expectation value of the Polyakov loop:
\begin{align}
    \avr{P_{\bs x}}_| (d) = \exp\bigl\{ - L_T F_{Q|}(d)\bigr\}\,,
    \label{eq_P_boundary}
\end{align}
placed at the point ${\bs x} = (x_1,x_2,d)$ in the presence of a single mirror (with fixed $x_3 =0$) and averaged over the tangential coordinates $x_1$ and $x_2$. In Eq.~\eq{eq_P_boundary}, $L_T$ is the lattice length in the imaginary time direction which also serves as an infrared regulator. At finite temperature $T$, the length $L_T$ is fixed, $L_T = 1/T$, and the term in the exponent of Eq.~\eq{eq_P_boundary} reduces to the familiar ratio $F/T$. 

In the thermodynamic limit at zero temperature, $L_T \to \infty$,  the Polyakov loop observable vanishes identically, making it practically impossible to calculate the potential~\eq{eq_P_boundary} of the heavy quark at large $L_T$. This property is of a kinematic rather than dynamical origin, shared by any (even unconfined) massive particle with a finite free energy $F > 0$. Therefore, to prove {\it qualitatively} the existence of an attractive interaction between a single quark and the mirror, we consider rather a small lattice with the temporal extension $L_T = 12 a$, in which the spatial correlator is limited to a few lattice steps due to finite-volume effects. 

The expectation value~\eq{eq_Polyakov_loop} contains unphysical distance-independent contributions, usually subtracted via a renormalization procedure. Due to the small volume of the lattice, it is challenging to renormalize the quark-free energy via its short-distance behavior, as it is usually done at finite temperature~\cite{Kaczmarek:2002mc}. We notice, however, that the free energy should flatten at the point $d = 6 a$ at the middle of the lattice due to the periodicity of the lattice. The flattening at this point is a $\beta$-independent feature, which we use as a renormalization requirement to calculate the renormalized free energy $F^{\mathrm{ren}}_{Q|}(l) = F_{Q|}(l,\beta) - F_0(\beta)$ near the mirror. The distance-independent subtraction term is described by remarkably simple linear dependence: $F_0(\beta) = -15.5 + 2.9 \beta$.

\begin{figure}[ht]
\centering
  \includegraphics[width=0.975\linewidth]{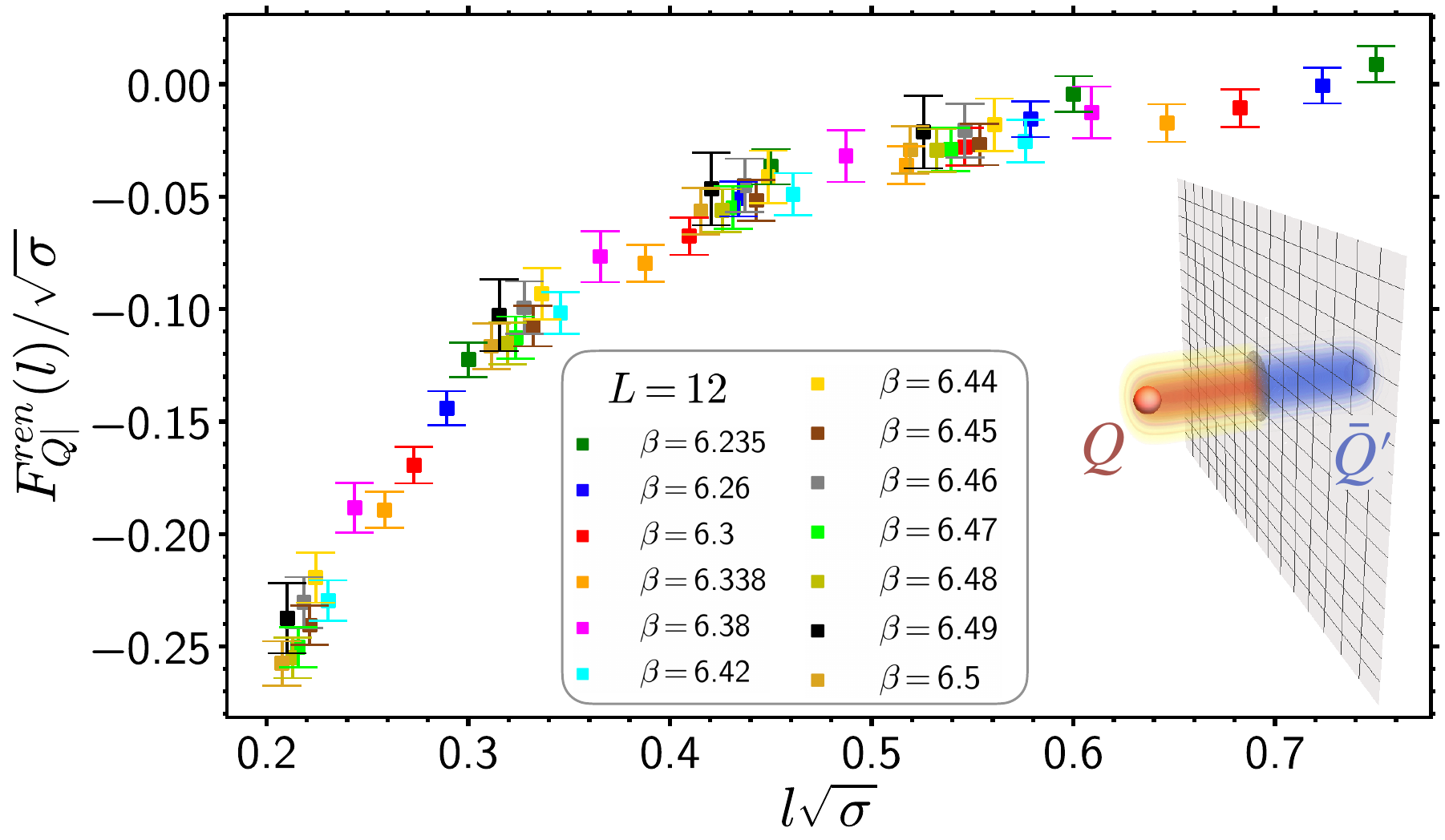}
  \caption{The renormalized free heavy-quark energy $F^{\mathrm{ren}}_{Q|}(l)$ at a distance $l$ from the chromometallic mirror, plotted in physical units, for various lattice coupling constants $\beta$ at $12^3$ lattice. The inset visualizes a quarkiton with the quark $Q$ and its negative image in the chromometallic mirror, the anti-quark $\bar Q'$, connected by a confining string (the ``mirror'' part of the string is shown in blue).}
  \label{fig_corr_wall}
\end{figure}

The renormalized free energy of heavy quark near the mirror, shown in Fig.~\ref{fig_corr_wall}, exhibits reasonable physical scaling because the points with different lattice cutoffs $a = a(\beta)$ collapse to the same smooth curve. We observe that the flat mirror attracts the quark along the normal direction, thus supporting the formation of the quarkiton bound state. The flattening of the free energy at larger distances $l$ is due to a finite volume effect which should disappear at larger volumes. At shorter distances, $F^{\mathrm{ren}}_{Q|}(l)$ shows qualitative signs of the expected linear behavior. Since the system resides far from the thermodynamic limit, all conclusions drawn from Fig.~\ref{fig_corr_wall} should be considered qualitative statements.

\vskip 1mm
\paragraph{\bf Quarkiton and color confinement.}

The color confinement property of the low-temperature (hadronic) phase requires that the asymptotic physical states of QCD must be colorless states of quarks and gluons. It is always concluded that the quark confinement implies that an isolated quark possesses infinite free energy and, therefore, cannot exist in the hadronic phase~\cite{Simonov:1996ati}. 

Strikingly similar physical properties are shared by fractional vortices in interacting multi-component Bose-Einstein condensates in two spatial dimensions as domain walls (strings in $2d$) linearly confine the vortices in bound states that resemble hadrons in QCD~\cite{Son:2001td}. Consequently, an isolated vortex, similarly to an isolated quark, cannot exist in the bulk of the condensate as a long domain wall attached to the vortex makes its energy infinite. However, single fractional vortices can still survive near the edge of the system, forming a bound state with its boundary (for fractionally charged vortices in superconductors with multi-band condensates and the boundary bound states, see Refs.~\cite{Silaev2011,Agterberg2014,Maiani2022}). 

\begin{figure}[ht]
\centering
  \includegraphics[width=0.925\linewidth]{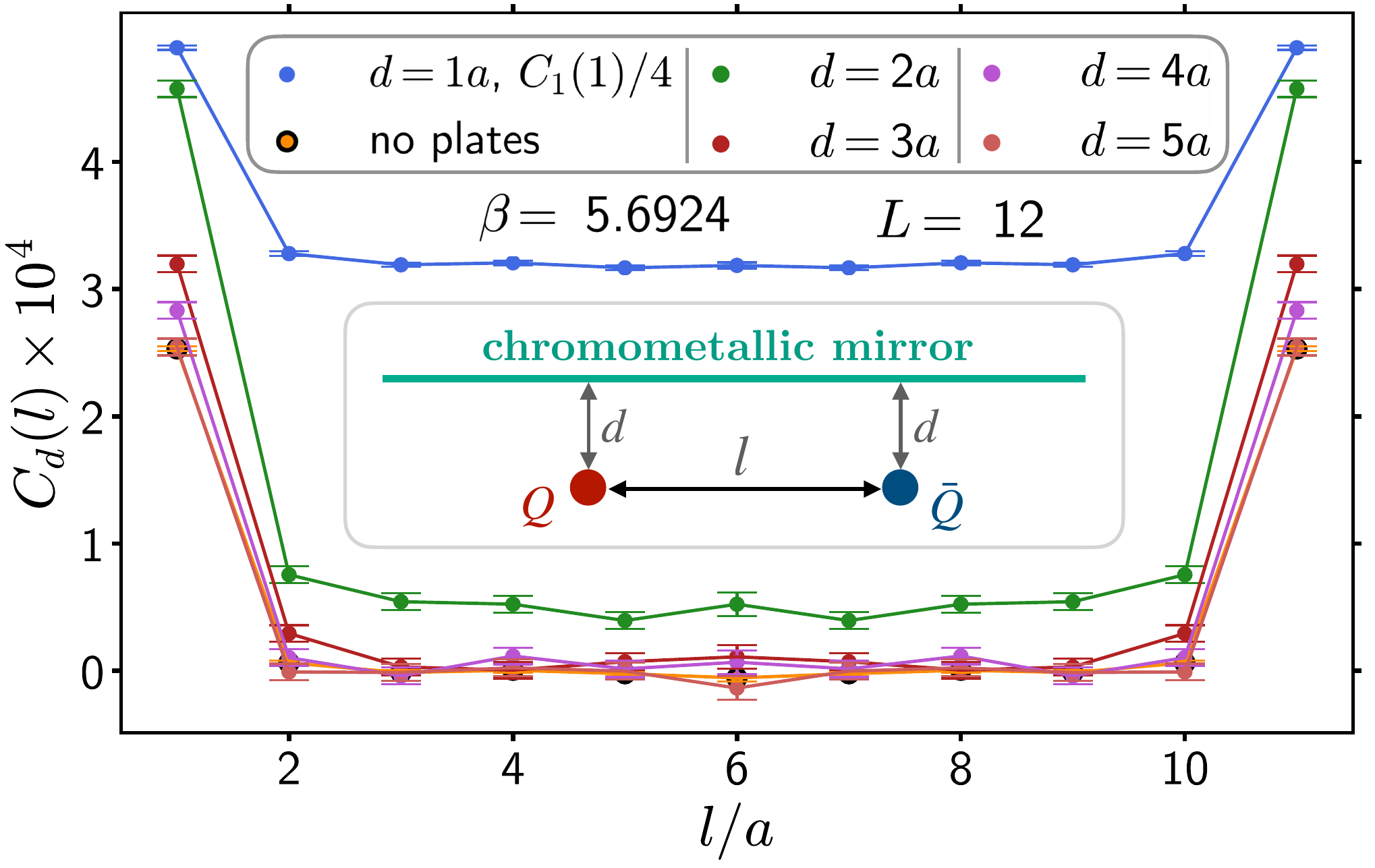}
  \caption{Correlator of the Polyakov loops $C_d(l)$ for a quark and an antiquark located at the fixed distance $d$ from the chromometallic mirror and separated by the distance $l$ from each other on the lattice $12^4$ at $\beta = 5.6924$ ($a \sqrt{\sigma} \simeq 0.4$~\cite{Athenodorou_2020ani}). The correlator at $d=1 a$ is scaled by the factor $1/4$. The correlator in the absence of the plates is also shown.}
  \label{fig_corr_ploops}
\end{figure}

\vskip 1mm
\paragraph{\bf Quarkiton interactions.} Our interpretation of the quarkiton boundary states can also be qualitatively supported by investigating the interactions of two quarkitons near the mirror. Let us consider a quark $Q$ and an antiquark $\bar Q$ located at the same distance $d$ from the boundary and at a distance $l$ as shown in the inset of Fig.~\ref{fig_corr_ploops}. Neglecting the short-distance Coulomb interaction via perturbative gluons, we consider the simplest confining string model, which implies that the energy of a mesonic, quark-antiquark $Q{\bar Q}$ state is $E_1 = \sigma l$. However, if $Q$ and $\bar Q$ form quarkiton states with, respectively, their mirror images $\bar Q'$ and $Q'$, then the total energy of this system is $E_2 = 2 E_{\mathrm{gl}} = 2 \sigma d$ (we neglect the interaction of the string with the mirror as well as perturbative gluonic exchanges). Therefore, energy arguments suggest that at short $Q{\bar Q}$ separation, $l < 2d$, the common mesonic $Q{\bar Q}$ state gets formed. As the separation increases at $l > 2d$, the string rearranges, and the meson decays into two quarkiton states,  $Q{\bar Q} \to  Q{\bar Q}' + {\bar Q}Q'$. 

The string rearrangement can be seen in Fig.~\ref{fig_corr_ploops}, although our relatively small ($L=12$) lattice does not allow us to observe it in detail. At large separation, $d$ from the mirror, $d=5a$, the correlator of Polyakov loops, $C_d(l) = \avr{P(x) P^*(x+l)}_d$ as the function of their mutual distance $l$, coincides with the same correlator in the absence of the mirrors. Thus, no quarkiton states are formed (a quark attracts to antiquark). As the distance to the mirror diminishes, the correlator increases in magnitude. At a small distance to the mirror $d$, the correlator reaches the plateaus in $l$, implying that the free energy of quarks does not depend on their separation $l$. This physical picture is perfectly consistent with the formation of the quarkiton: the quark and the antiquark attract to their images in the mirror. Moreover, again expectedly, the plateau at $d=1a$ is higher than at $d=2a$, in agreement with the fact that the string between the (anti)quark and its image in the mirror is shorter for the quark which is located closer to the mirror. At short distances, the perturbative Coulomb interaction prevails over the string effects~\cite{Brambilla_1999ja}, but this fact does not change our conclusions given the monotonic nature of the attractive $Q\bar Q$ potential. 

The boundary (glueton and quarkiton) states can also interact with the bulk (glueball and meson) states. Also, two quarkiton states, confined to the boundary, can combine by producing a colorless meson state which can then propagate into the bulk of the system.

\vskip 1mm
\paragraph{\bf A heavy quark between non-Abelian mirrors.}

Finally, we address the nature of the vacuum between two chromometallic mirrors. In Ref.~\cite{Chernodub_2018pmt}, the same question has been raised in two spatial dimensions for the vacuum of SU(2) gauge theory in between two parallel wires (plates in Euclidean spacetime). It was concluded that in the confining low-temperature phase, the approaching plates generate a deconfinement phase in the space between them. The deconfinement mechanism in this non-Abelian theory has been related to an identical effect in the 2+1 dimensional compact Abelian gauge model~\cite{Chernodub_2017gwe} where the Casimir-induced deconfinement can be explained analytically~\cite{Polyakov:1976fu}.

We see no pronounced signatures of a phase transition in the space between the plates in the behavior of the non-Abelian Casimir energy~\eq{eq_Cas_Lattice} shown in Fig.~\ref{fig_Casimir_energy}. To quantify the effect of the chromometallic mirrors on (de)confining properties of the vacuum, we study another quantity, the unrenormalized free energy of a heavy quark calculated in the space between the plates:
\beqn
L_T F_{Q}^{\mathrm{Cas}}(R) = - \ln |P|_{V(R)} \equiv - \ln \Bigl\langle \Bigl| \sum_{{\bs x} \in V(R)} P_{\bs x} \Bigr| \Bigr\rangle, \quad
\label{eq_F_Q}
\eeqn
where the expectation value of the Polyakov loop $|P|_V$ is taken only over the volume $V = V(R)$ between the mirrors separated by the distance $R$.

\begin{figure}[ht]
\centering
  \includegraphics[width=0.975\linewidth]{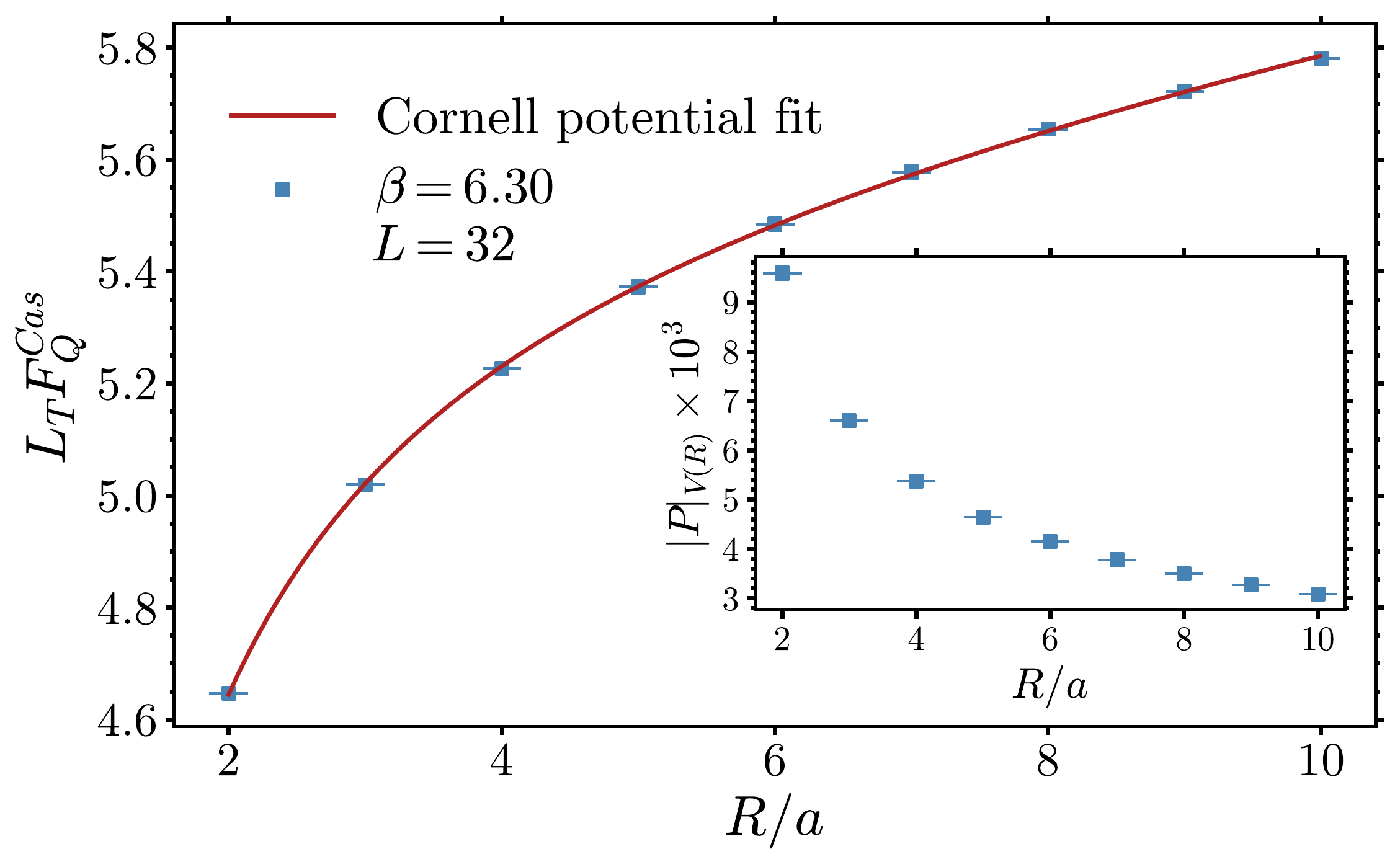}
  \caption{Mean free energy of heavy quark in between the mirrors~\eq{eq_F_Q} as function of the inter-mirror separation $R/a$ (in lattice units) on the lattice $32^4$. The red line is the best fit by the Cornell potential~\eq{eq_scaling_lat} with the fit parameters $c_1=2.03(4)$, $c_2=0.044(1)$, and $c_0=5.55(2)$. The inset shows the expectation value of the corresponding Polyakov loop~\eq{eq_F_Q}.}
  \label{fig_freeenergypol_latdist}
\end{figure}

In the inset of Fig.~\ref{fig_freeenergypol_latdist}, we show the Polyakov loop in between the plates $|P|_{V(R)}$. This quantity takes a finite value at small inter-plate separations $R$ and then quickly diminishes with increasing distance between the plates. Such behavior points to an effective deconfinement regime between the closely-spaced plates, which we interpret as a signal of the formation of (a superposition of) finite-energy quarkiton states between the test quark and its antiquark image in the mirrors. As the distance between the plates increases, the free energy of long-stretched quarkiton states rises, and the Polyakov loop vanishes, thus signaling the onset of the confining regime. 

The phenomenological interaction between quarks and antiquarks is often described by a Cornell-type potential that combines a linear string behavior at long distances with a short-distance Coulomb interaction~\cite{Brambilla_1999ja}. Therefore, the mean free energy of a quarkiton, in which a quark interacts with its antiquark image in the mirror, should follow a similar behavior with the typical quarkiton size set by the inter-plate separation $R$. This phenomenological expectation is indeed confirmed in Fig.~\ref{fig_freeenergypol_latdist}, showing that the free energy~\eq{eq_F_Q} is indeed excellently described by the Cornell potential:
\beqn
L_T F_{\mathrm{Q}}^{\mathrm{Cas}}(R/a)=-\frac{c_1}{R/a} + c_2 \frac{R}{a} + c_0\,,
\label{eq_scaling_lat}
\eeqn
where $c_a$ (with $a = 0,1,2$) are the fitting parameters. Equation~\eq{eq_scaling_lat} implies that at short inter-plate separations, a heavy quark in the space between the mirrors possesses a finite free energy which we interpret as a deconfinement of color. As the distance between the plates $R$ increases, the free energy increases, leading to the exponential vanishing of the Polyakov loop and the onset of the color confinement.

\vskip 1mm 
\paragraph{\bf Conclusions.} Using first-principle numerical simulations, we calculated the nonperturbative non-Abelian Casimir energy generated by two closely spaced chromometallic mirror plates in SU(3) Yang-Mills theory. We also revealed the presence of a new gluonic excitation, the glueton, which we interpret as a colorless bound state of a gluon with its image in a chromometallic mirror. 

The glueton is a non-topological excitation that shares similarities with a surface exciton in a superconductor. Unexpectedly, the glueton mass~\eq{eq_mass_glueton} turns out to be lower than the mass of the groundstate $0^{++}$ glueball. This property of the glueton (``the edge mode is lighter than the mass gap in the bulk'') is shared by its topological analogs in the condensed matter such edge modes in topological insulators~\cite{Konig_2008} or the Volkov-Pankratov states at the interfaces of semiconductors~\cite{volkov1985two}.

The presence of boundaries also affects the dynamics of quarks. We show that similarly to confined fractional vortices in multi-component condensates~\cite{Son:2001td,Silaev2011,Agterberg2014,Maiani2022}, a single isolated quark can exist in the hadronic phase of QCD near (and confined to) a large perfect chromometallic mirror, forming a colorless boundary state: the quarkiton. The glueton and quarkiton states can be relevant near domain walls in QCD and QCD-like theories.

\vskip 1mm
\begin{acknowledgments}
\paragraph{\bf Acknowledgments.}
The authors are grateful to Julien Garaud for making them aware about Refs.~\cite{Silaev2011,Agterberg2014,Maiani2022}.
The numerical simulations were performed at the computing cluster Vostok-1 of Far Eastern Federal University.
The work of VAG, AVM, and AST was supported by Grant No. 0657-2020-0015 of the Ministry of Science and Higher Education of Russia.
\end{acknowledgments}

\bibliography{casimirSU3}

\end{document}